\documentclass[doublecol]{epl2} 
\usepackage{gensymb}

\title{Subwavelength sound screening by coupling space-coiled Fabry-Perot resonators}
\shorttitle{Subwavelength sound screening} %Insert here a short version of the title if it exceeds 70 characters

\author{A. Elayouch \inst{1} \and M. Addouche \inst{1} \and M. Farhat \inst{2}  \and A. Khelif \inst{1}}

\institute{
	\inst{1} Institut FEMTO-ST, CNRS, Universit\'e de Bourgogne Franche-Comt\'e, 15B avenue des Montboucons, 25030 Besancon Cedex, France\\
	
	\inst{2} Qatar Environment and Energy Research Institute (QEERI), Hamad Bin Khalifa University, Qatar Foundation, Doha, Qatar
	
}

\pacs{43}{Acoustics}

\abstract{
We explore broadband and omnidirectional low frequency sound screening based on locally resonant acoustic metamaterials. We show that the coupling of different resonant modes supported by Fabry-Perot cavities can efficiently generate asymmetric lineshapes in the transmission spectrum, leading to a broadband sound opacity. The Fabry-Perot cavities are space-coiled in order to shift the resonant modes under the diffraction edge, which guaranty the opacity band for all incident angles. Indeed, the deep subwavelength feature of the cavities leads to avoid diffraction that have been proved to be the main limitation of omnidirectional capabilities of locally resonant perforated plates. We experimentally reach an attenuation of few tens of dB at low frequency, with a metamaterial thickness fifteen times smaller than the wavelength ($\lambda / 15$). The proposed design can be considered as a new building block for acoustic metasurfaces having a high level of manipulation of acoustic waves.}

\begin{document}

\maketitle

The advent of phononic crystals in the landscape of elastic/acoustic waves has opened up new prospects in understanding and controlling wave propagation~\cite{cummer_controlling_2016, lee_acoustic_2017, khelif_acoustic_2010}. Indeed, phononic crystals use fundamental properties of waves, such as diffusion or Bragg interference phenomena, in order to generate the so-called band gaps, which are basically frequency bands where there is no wave propagation ~\cite{wilm_out--plane_2003}. Sound control, in the audible range, is the most evident application of band gaps. For the human ear, the frequency range under consideration is approximately from 20 Hz to 20 kHz, which corresponds to wavelength up to several meters in air. Therefore, dealing with a sonic crystal with a period in this size range would be certainly cumbersome. The sculpture \'{O}rgano of Eusebio Sempere is an excellent example of this~\cite{martinezN1995}.

Recently, various studies involving resonators have shown how to overcome these limitations and reach promising applications ~\cite{khelif_locally_2010, addouche_superlensing_2014}. This is the case of the Locally Resonant Sonic Crystal (LRSC) introduced in 2000 by Liu \emph{et al.} consisting of metal spheres coated with an elastic material, and for which a negative dynamic mass density is generated leading to band gaps at frequencies lower than that of a classical Bragg-based sonic crystal of equal dimensions~\cite{LiuScience2000}.  More recently, a panel version of the LRSC was proposed in the form of an elastic membrane fixed by a grid, in the middle of which was attached a small mass~\cite{yang_membrane-type_2008}. In this metamaterial, a near-total reflection occurs around the resonance. Besides these, recent studies have highlighted some properties arising from using space-coiling in acoustic metamaterials, introduced by Liang and Li ~\cite{liang_extreme_2012}. Among them, the possibility of generating extreme effective parameters, which offer promising applications such as negative refraction. Therefore, subsequent studies have brought to light both non-resonant ~\cite{frenzel_three-dimensional_2013, liang_space-coiling_2013, xie_measurement_2013, xie_tapered_2013}, and resonant~\cite{song_emission_2014} acoustic metamaterials involving space-coiling. The latter ones were used to improve sensors capacities of detection by generating a sound pressure level gain, even going so far as to achieve the Enhanced Acoustic Transmission for wavelength 15 times bigger than the periodicity ~\cite{li_extraordinary_2013}. In this sense, space-coiling represents an interesting alternative for their capacity to target greater wavelengths. However, the relative bandwidth of this kind of attenuation is fairly narrow, as is often the case in resonant phenomena. Indeed, for relatively simple systems, these resonances remain classical, in the sense that they exhibit symmetric Lorentzian lineshapes ~\cite{ park_giant_2013, lu_extraordinary_2007}. But since this past century, resonances with asymmetrical lineshapes have been highlighted: they are referred to as Fano resonances~\cite{miroshnichenko_fano_2010}. In fact, Ugo Fano explained them as originating from the constructive and destructive interference of a narrow discrete resonance with a broad spectral line or continuum ~\cite{Fano61, Christian2013}. Since then, numerous studies have allowed a better understanding of Fano resonances in various fields of physics such as electromagnetism, plasmonics~\cite{liu_plasmonic_2009}, and more recently acoustics~\cite{Liu10,Santillan11, amin_acoustically_2015}, through the acoustical analogue of Electromagnetically Induced Transparency (EIT) ~\cite{fleischhauer_electromagnetically_2005}. Recently, the opaque counterpart of Acoustically Induced Transparency (AIT) were also achieved by using an array of Fabry-Perot (FP) resonators in the ultrasonic regime. Such resonators were also coupled to each others, except that the coupling was carried out laterally by the fluid-solid interaction~\cite{elayouch_experimental_2013}. Such a coupling has proven to be efficient at normal incidence, but more limited at oblique incidence. Indeed, apertures when forming a periodic array also act as a diffraction grating~\cite{Diffwiley2007}. In this regard, it was shown that the main limitation of an omnidirectional screening based on such resonators was directly linked to diffraction~\cite{elayouch_how_2016}. 

In this paper, we investigate the transmission properties of an acoustic metamaterial constituted of space-coiled cavities and operating at low frequency compared to the thickness. For this, we design and fabricate a structure made of Fabry-Perot cavities by using a three-dimensional (3D) printer. An asymmetric Fano lineshape of transmission can take place when two modes of neighboring cavities interfere~\cite{amin_acoustically_2015}. In the proposed design, this is achieved by considering Fabry-Perot cavities having different quality factors and by placing them side by side. We experimentally show that a wide frequency band of sound opacity can be found through the proposed acoustic metamaterial with subwavelength thickness. Additionally, we numerically demonstrate that this low frequency screening behavior avoids the frequency of diffraction occurrence, which has been proved to be the main limitation of the omnidirectional capabilities of locally resonant perforated plates.

\begin{figure}[!h]
	\centerline{\includegraphics[width=8.6cm,keepaspectratio]{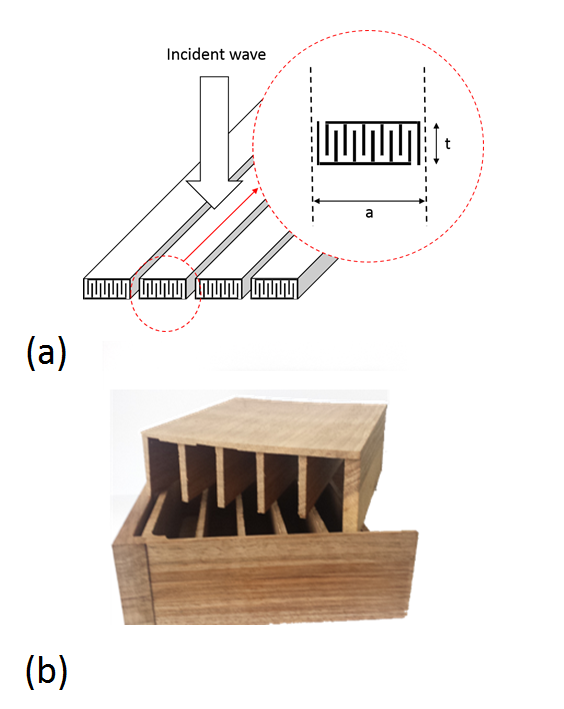}}
	\caption{(a) Schematics of the periodically perforated plate, and zoom of a unit cell constituting the array, and made of two Fabry-Perot cavities, one of which is space-coiled. The metamaterial thickness and the period are : $t=50$ mm, $a=30$ mm. The cavities are $7$ mm-wide, and the wall thickness is $2$ mm. The metamaterial is excited with a sound wave traveling in the $y$-direction (from bottom to top).
		(b) Photograph of the acoustic metamaterial fabricated by 3D printing. The sample is constituted by an array of 3 cells, where each unit cell is made of two Fabry-Perot cavities, one of which is space-coiled.  The metamaterial thickness is equal to $t=50$ mm, the system period $a=30$ mm, and the aperture width is $d=7$ mm.}
	\label{fig:fig00}
\end{figure}

Fabry-Perot resonances are well-known phenomena in optics, particularly for lasers where the amplification involves light rays confinement in resonant cavities. In what follows, we first lay the bricks which represent the basic elements of our acoustic study, by introducing the elemental structures that we work on, as well as their limitations. For this, we first consider the case of a plate periodically constituted of slits, as represented in Figure~\ref{fig:fig00}a, and we begin by considering interferences that can occur when dealing with such structures. 

%%%%%%%%%%%%%%%%%%%%%%%%%%%%%%%%%%%%%%%%%%%%%%%%%%%%%%%%%%%%%%%%%%%%%%%%%%%%%%%%%%%%%%%%%%%%%%%%%%%

\begin{figure}[!h]
	\centerline{\includegraphics[width=8.6cm,keepaspectratio]{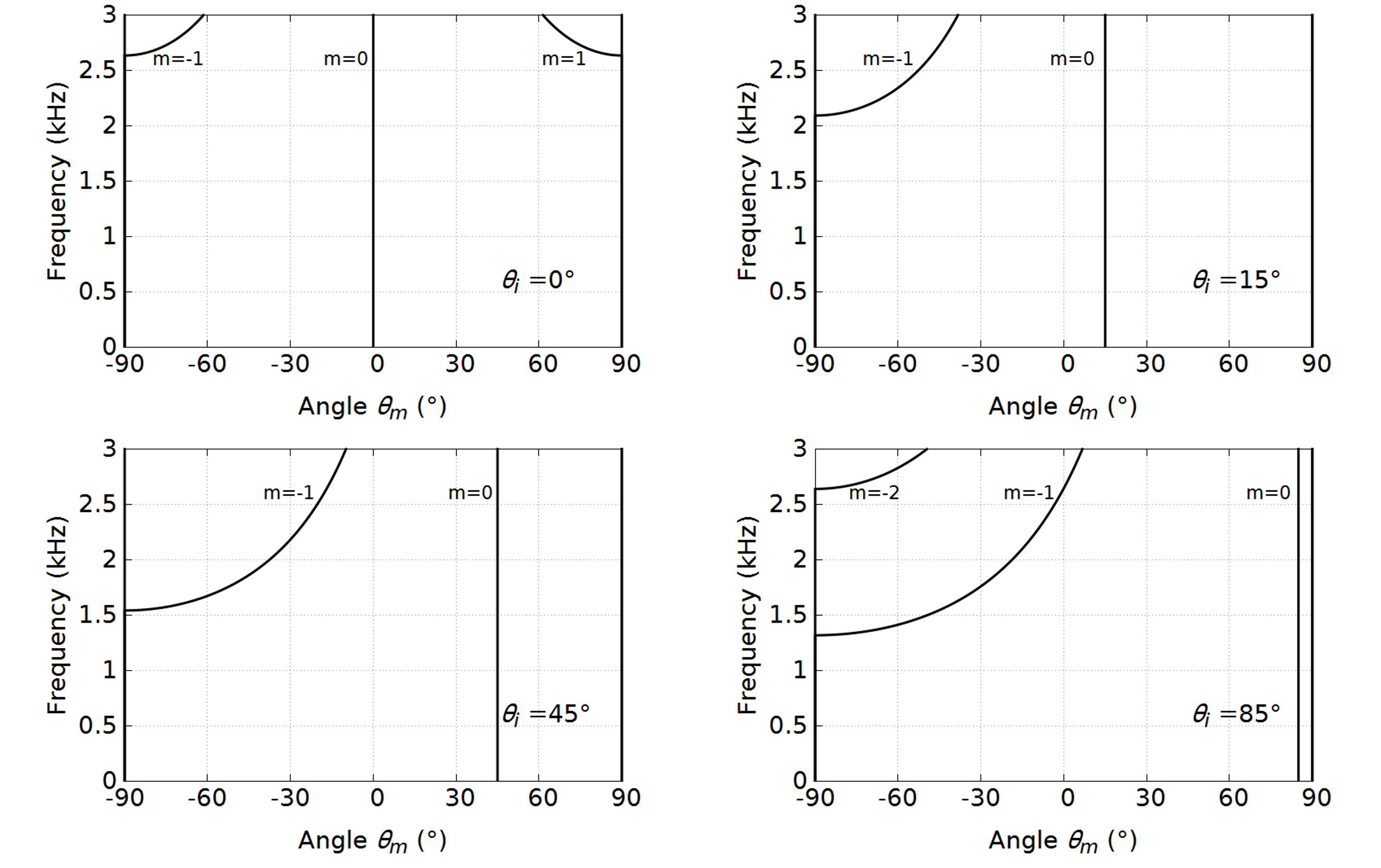}}
	\caption{Frequency localization $\theta_{m}$ of the first orders of diffraction $m=-1$ and $0$ in the case of different incident angles $\theta_{i}=0\degree$, $\theta_{i}=15\degree$, $\theta_{i}=45\degree$, and $\theta_{i}=85\degree$. The frequency of diffraction occurrence is obtained analytically from the grating law $\theta_{m}=\arcsin (\frac{m\lambda}{a} +sin\theta_{i})$, as function of the angles of incidence.}
	\label{fig:fig2}
\end{figure}

Indeed, periodic apertures act as diffractive elements, just as the well-known Young's slits in optics. Thus, this structure constituted of slits array and considered as a diffraction grating that obeys the following law~\cite{royerBOOK1999}:
\begin{center}$a\cdot (sin(\theta_{m})-sin(\theta_{i}))=m\cdot\lambda/n$\end{center}
Where \textit{i} stands for incidence, \textit{a} is the slit array period, \textit{m} is the order of diffraction, $\lambda$ is the wavelength, and \textit{n} the refractive index. 
Thus, for an incidence angle of $0\degree$, $15\degree$, $45\degree$, and $85\degree$, and by considering only the $m=-1$ and $m=0$ orders of diffraction, we obtain analytically the curves of Figure~\ref{fig:fig2}, which provides the frequency localization of these orders of diffraction. It is possible to predict the frequency where diffraction occurs for such a structure only based on its geometrical features. Thus, for a period of $a=129$ mm, the frequency marking the limit is approximately $1320$ Hz, from which $m=-1$ order of diffraction occurs. It has been proved in the case of ultrasonic waves~\cite{elayouch_how_2016}, that diffraction may prevent the generation of acoustic reflection based on resonances. Hence, in order to generate a resonant reflection mechanism, it might be appropriate to consider the low frequency regime, and so to remain well below the identified frequency of diffraction occurrence.

%%%%%%%%%%%%%%%%%%%%%%%%%%%%%%%%%%%%%%%%%%%%%%%%%%%%%%%%%%%%%%%%%%%%%%%%%%%%%%%%%%%%%%%%%%%%%%%%%%%
%%%%%%%%%%%%%%%%%%%%%%%%%%%%%%%%%%%%%%%%%%%%%%%%%%%%%%%%%%%%%%%%%%%%%%%%%%%%%%%%%%%%%%%%%%%%%%%%%%%

For this, we perform a numerical study using the finite element method with COMSOL Multiphysics~\cite{_comsol_2008}, and take advantage of the periodicity by applying periodic conditions. We study the propagation of acoustic waves in air and consider rigid conditions at the interface between the solid and the fluid. We are specifically interested in obtaining the acoustic transmission properties at normal incidence, in the first instance. In the fluid, sound waves are governed by the following equation for the differential pressure $p$
\begin{equation}
\nabla\cdot\left(\rho^{-1}\nabla p\right)+\frac{\omega^2}{\rho c^2}p=0\, ,
\label{eq:eq0}
\end{equation}
where $\rho$ is the mass density of the fluid, $c$ is the sound speed in air, and $\omega$ is the angular frequency of the acoustic pressure wave.
An acoustic source is positioned upstream of the structure and we calculate the pressure field  downstream to it, which permits us to ultimately obtain the transmission spectrum for yhe frequencies of interest.

\begin{figure}[!h]
	\centerline{\includegraphics[width=8.6cm,keepaspectratio]{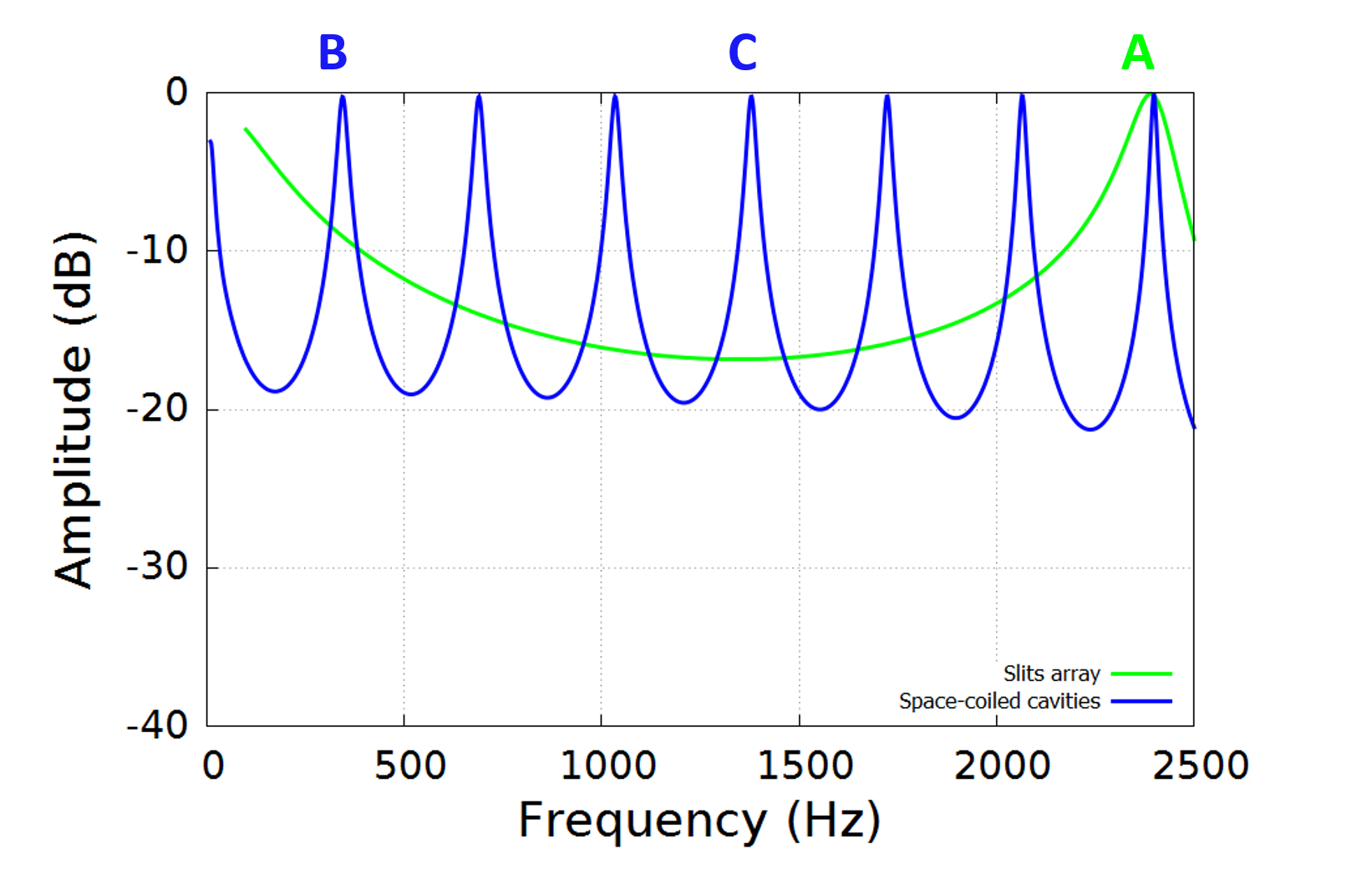}}
	\caption{The green and the blue lines represent the amplitude transmission spectra, obtained by finite element method, related to a plate perforated with periodically distributed Fabry-Perot cavities. The green line corresponds to the acoustic response of a slits array, and the blue line to an array of space-coiled cavities, having a period of $a=129$ mm. The aperture width is $d=7$ mm. The transmission curves are marked by letters A, B, and C, at the resonance frequencies $f_{A}=2390$ Hz, $f_{B}=345$ Hz, and $f_{C}=1380$ Hz.}
	\label{fig:fig1}
\end{figure}

\begin{figure}[!h]
	\centerline{\includegraphics[width=8.6cm,keepaspectratio]{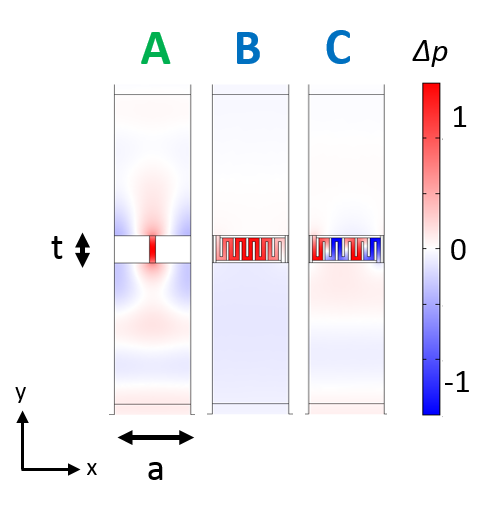}}
	\caption{Pressure fields, in and around the cavity, at the resonance frequencies $f_{A}=2390$ Hz, $f_{B}=345$ Hz, and $f_{C}=1380$ Hz, marked by the corresponding letters A, B, and C. The color scales have been normalized with normalization constants 10, 13.2, and 16.1, respectively.}
	\label{fig:fig1a}
\end{figure}

We first consider the case of a plate periodically constituted of straight slits, wherein each of the slits plays the role of a Fabry-Perot resonant cavity. At the resonance frequency of this straight cavity, it is possible to observe the effect of enhanced acoustic transmission when an acoustic wave impinges such a plate at normal incidence. Figure~\ref{fig:fig1} represents the transmission spectrum of such a plate having a thickness $t=50$ mm, constituted of $7$ mm-wide slits arranged with a period of $a=129$ mm. As can be clearly seen in Figure~\ref{fig:fig1}, a resonance appears at the frequency of $f_{A}=2390$ Hz, as a matter of fact, $f=c/2t$, which corresponds to a wavelength $\lambda$ twice bigger than the plate thickness. In Figure~\ref{fig:fig1}, this transmission profile is marked by letter A, and the pressure field, in and around the cavity, is represented at this resonance frequency in Figure~\ref{fig:fig1a}. As mentioned previously, and due to the resonant character of the cavity, the acoustic energy confinement can be clearly observed through the large variation of the acoustic pressure in the slit.   

%%%%%%%%%%%%%%%%%%%%%%%%%%%%%%%%%%%%%%%%%%%%%%%%%%%%%%%%%%%%%%%%%%%%%%%%%%%%%%%%%%%%%%%%%%%%%%%%%%%
%%%%%%%%%%%%%%%%%%%%%%%%%%%%%%%%%%%%%%%%%%%%%%%%%%%%%%%%%%%%%%%%%%%%%%%%%%%%%%%%%%%%%%%%%%%%%%%%%%%

%%%%%%%%%%%%%%%%%%%%%%%%%%%%%%%%%%%%%%%%%%%%%%%%%%%%%%%%%%%%%%%%%%%%%%%%%%%%%%%%%%%%%%%%%%%%%%%%%%%
%%%%%%%%%%%%%%%%%%%%%%%%%%%%%%%%%%%%%%%%%%%%%%%%%%%%%%%%%%%%%%%%%%%%%%%%%%%%%%%%%%%%%%%%%%%%%%%%%%%

As said before,  we want to remain well below the identified frequency of diffraction occurrence in order to generate a resonance-based reflection mechanism. We therefore consider the case of Fabry-Perot cavities having the particularity to be space-coiled. Such cavities permit the generation of resonances for wavelength much larger than what the plate thickness leads us to believe. Figure~\ref{fig:fig1} represents the transmission spectrum for the case of a plate having a thickness of $t=50$ mm, but constituted of  $7$ mm-wide space-coiled cavities, and arranged with a period of $a=129$ mm, as represented in Figure~\ref{fig:fig1a}. We can clearly observe the occurrence of a series of Fabry-Perot resonances from a frequency as low as $f_{B}=345$ Hz. At this frequency, the wavelength $\lambda$ equals $1$ m, which is almost $20$ times larger than the plate thickness. The equivalent straight Fabry-Perot cavity should be long of $l=c/2f= 500$ mm.
The transmission profile is marked by letters B and C in Figure~\ref{fig:fig1}, at the frequencies $f_{B}=345$ Hz and $f_{C}=1380$ Hz. For the corresponding frequencies, we also represented the pressure fields in Figure~\ref{fig:fig1a}. We can particularly observe the acoustic pressure variations in the cavity, which correspond to a half-wavelength for the fundamental frequency marked by letter B, and which correspond to 4 half-wavelength for the fourth harmonic marked by letter C.   

In order to gain a better understanding of the physical mechanisms occurring in space-coiled cavities, we also study the propagation of acoustic waves in the direction of the periodicity. For this, we first calculate the band structure for a system consisting of Fabry-Perot-like cavities, taking the shape of straight slits and space-coiled ones, embedded in a rigid matrix.  
In order to achieve this, we calculate the dispersion curves by using a finite element method. The grating, assumed as an infinite system, is modeled as a unit cell for which Bloch-Floquet conditions are implemented via periodic boundary conditions in the x direction. This comes down to apply a phase relation on the lateral sides of the mesh, defining boundary conditions between adjacent cells. This phase relation is related to the Bloch wavenumber of the modes of the periodic structure. We vary the wave vector in the first Brillouin zone and solve an eigenvalue problem, which permits us to obtain the eigenfrequencies~\cite{khelifBOOK2015}. 
In addition, we calculate the transmission spectra for the cases of acoustic waves impinging the grating at different incidence angles varying from $0$ to $90$ degrees. This permits us to complete our dispersion curves in the radiative zone of the band structure, by representing the transmission obtained as function of the In-Plane Wavevector $k_{x} = k_{0} sin(\theta_{i})$. 

%%%%%%%%%%%%%%%%%%%%%%%%%%%%%%%%%%%%%%%%%%%%%%%%%%%%%%%%%%%%%%%%%%%%%%%%%%%%%%%%%%%%%%%%%%%%%%%%%%%
\begin{figure}[!h]
	\centerline{\includegraphics[width=8.6cm,keepaspectratio]{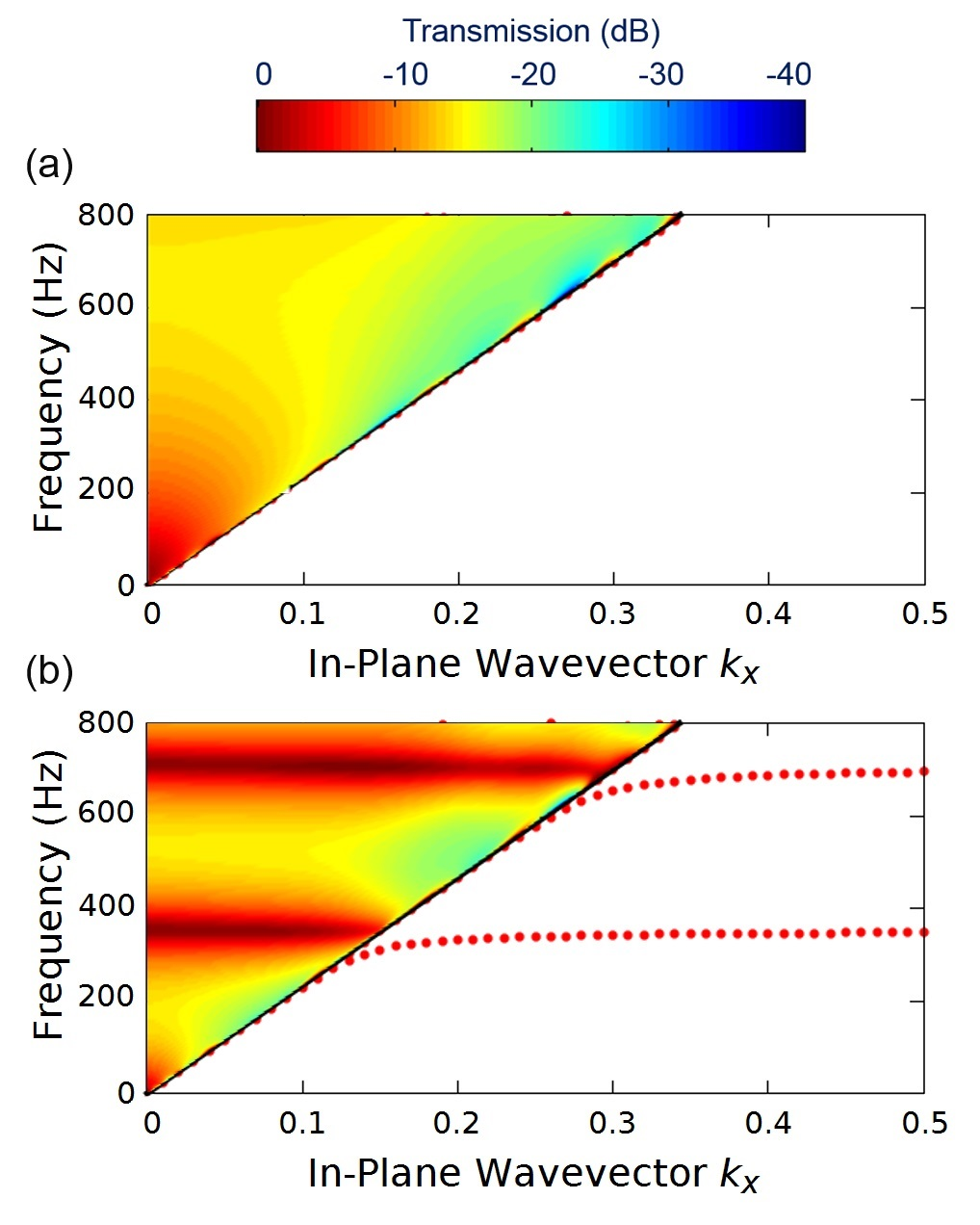}}
	\caption{Amplitude transmission spectra function of the In-Plane Wavevector $k_{x} = k_{0} sin(\theta_{i})$, and dispersion obtained numerically by finite element method. The acoustic metamaterial constituted by an array of resonators, where each unit cell is made of two Fabry-Perot cavities, one of which is space-coiled.  The metamaterial thickness is equal to $t=50$ mm, the system period $a=129$ mm, and the aperture width is $d=7$ mm.}
	\label{fig:fig5a}
\end{figure}
%%%%%%%%%%%%%%%%%%%%%%%%%%%%%%%%%%%%%%%%%%%%%%%%%%%%%%%%%%%%%%%%%%%%%%%%%%%%%%%%%%%%%%%%%%%%%%%%%%%

Figure~\ref{fig:fig5a} shows the dispersion, for the unit cells shown in Figure~\ref{fig:fig1a}. The dispersion is calculated up to the frequency of 800 Hz for both configurations. In the first configuration corresponding to simple slit arrays (see Figure~\ref{fig:fig5a}a), we simply observe the presence of solutions in the limit of the sound cone, that marks the boundary of the radiative zone. By contrast, the grating made of space-coiled cavities exhibits two modes (see Figure~\ref{fig:fig5a}b), which are materialized by two acoustic flat bands. Such modes are therefore strongly localized in the slit cavity with FP-like fields, where the cavity's length and width are the dominant variables in determining the resonance position.

In the radiative zone, we also observe resonant modes for the grating made of space-coiled cavities, in contrary to straight slit grating, in the [0-800] Hz frequency range. These resonant modes can be excited using an acoustic source positioned in the far field. They are characterized by flat bands of near-total transmission that do not depend on the In-Plane Wavevector $k_x$, except near the sound line where we can observe slight deviations. Such deviations can be linked to the hybridization with surface acoustic waves. Such an hybridization is clearly observable in the non-radiative zone near the sound line.

%%%%%%%%%%%%%%%%%%%%%%%%%%%%%%%%%%%%%%%%%%%%%%%%%%%%%%%%%%%%%%%%%%%%%%%%%%%%%%%%%%%%%%%%%%%%%%%%%%%%
% Coupling
%%%%%%%%%%%%%%%%%%%%%%%%%%%%%%%%%%%%%%%%%%%%%%%%%%%%%%%%%%%%%%%%%%%%%%%%%%%%%%%%%%%%%%%%%%%%%%%%%%%%

We now numerically study the coupling effect between the resonant cavities previously introduced, through finite elements calculation. This permits us to obtain the pressure field in the light of transmission spectra obtained in response of an acoustic wave impinging our metamaterial. We therefore consider the case of a plate having a thickness of $t=50$ mm, and constituted of apertures arranged in a 1D array. Each period, $a=129$ mm-wide, is constituted of two resonant cavities. While the first cavity is a straight aperture, the second is a space-coiled one, as represented in Figure~\ref{fig:fig00}a and b. Thus, both are Fabry-Perot cavities having different resonance frequencies and quality factors. Besides, the cavities inputs and outputs are placed side by side in order to foster the interaction between them,  specifically through near-field coupling, as will be shown below. 

%%%%%%%%%%%%%%%%%%%%%%%%%%%%%%%%%%%%%%%%%%%%%%%%%%%%%%%%%%%%%%%%%%%%%%%%%%%%%%%%%%%%%%%%%%%%%%%%%%%
\begin{figure}[!h]
	\centerline{\includegraphics[width=8.6cm,keepaspectratio]{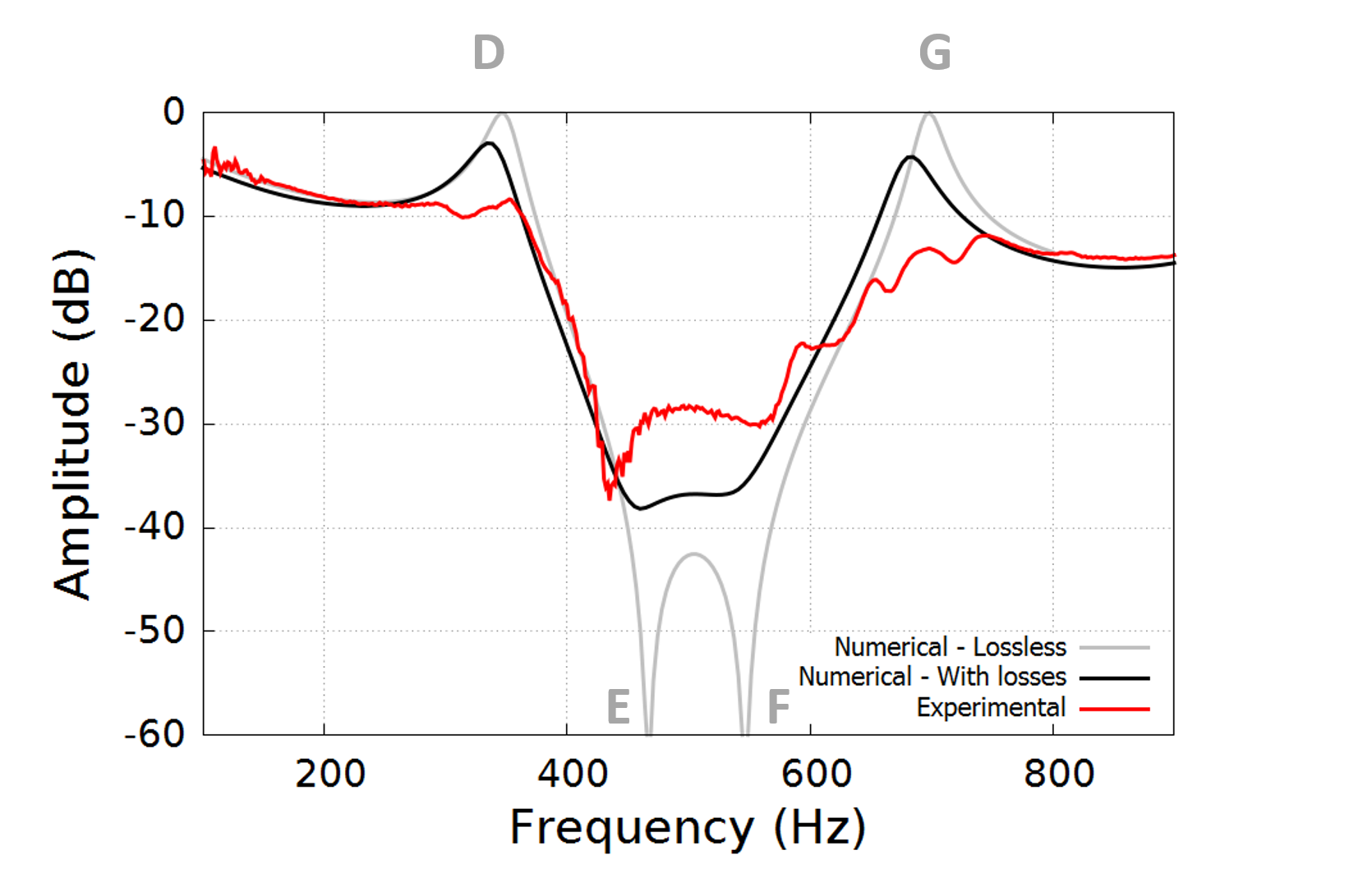}} % [width=0.9\columnwidth,height=0.5\columnwidth]
	\caption{Amplitude transmission spectra of the acoustic metamaterial, obtained numerically by finite element method, without (grey curve) and with losses (black curve) taken into account. (red curve) Amplitude transmission spectrum of the acoustic metamaterial obtained experimentally using a Kundt's tube.}
	\label{fig:fig8}
\end{figure}
%%%%%%%%%%%%%%%%%%%%%%%%%%%%%%%%%%%%%%%%%%%%%%%%%%%%%%%%%%%%%%%%%%%%%%%%%%%%%%%%%%%%%%%%%%%%%%%%%%%

The transmission spectrum at normal incidence is represented in Figure~\ref{fig:fig8}. We clearly observe the occurrence of a series of resonances and antiresonances leading to an attenuation band centered at $500$ Hz, with a transmission reaching $35$ dB of attenuation over a relative band of $30 \%$. The Fabry-Perot resonances with a symmetric profile, namely with Lorentzian profile in their classical expression, have been replaced by asymmetrical resonance profiles, which relates to the idea of a Fano-like interaction. Indeed, one of the main features of Fano resonances are their asymmetric line profiles where constructive interference corresponds to resonant enhancement and destructive interference to resonant suppression of the transmission. This is the case of the resonance marked by letter A in Figure~\ref{fig:fig1}, which has given rise, after coupling, to a resonance (marked by letter D) followed by an antiresonance (marked by letter E), as represented in Figure~\ref{fig:fig8}.

\begin{figure}[!h]
	\centerline{\includegraphics[width=8.6cm,keepaspectratio]{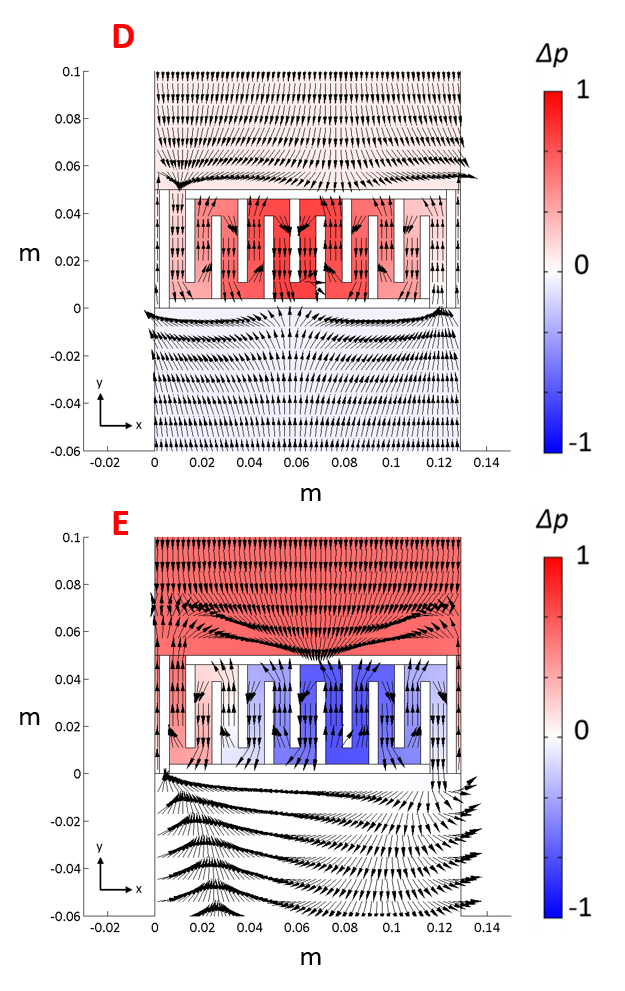}}
	\caption{Pressure and acoustic velocity vector fields, in and around the cavity, at the resonance frequency $f_{D}=345$ Hz, and the anti-resonance frequency $f_{E}=470$ Hz.}
	\label{fig:fig5}
\end{figure}

To better understand this interaction, we represent in Figure~\ref{fig:fig5} the pressure fields at the first resonance and anti-resonance frequencies. As mentioned previously, we observe the confinement of acoustic energy in both cavities. Indeed, the pressure level significantly increases at resonances. Here one can see as well that the pressure field in the transmission domain is almost zero at the anti-resonances (E), which confirms the near-total reflection of the acoustic waves.    
Moreover, it is interesting to visualize how the coupling of neighbor Fabry-Perot cavities generates such an interaction. This can bee seen in Figure~\ref{fig:fig5} where we can particularly observe the acoustic velocity vector fields at the frequencies of resonances and antiresonances marked by D and E. One may note that, for D, the vectors at the cavities inputs are in phase with the incident wave, and contribute together to near-total transmission (same as G). Near-total reflection is observer at E (same as F) when they are out of phase with the incident wave.

%%%%%%%%%%%%%%%%%%%%%%%%%%%%%%%%%%%%%%%%%%%%%%%%%%%%%%%%%%%%%%%%%%%%%%%%%%%%%%%%%%%%%%%%%%%%%%%%%%%
%%%%%%%%%%%%%%%%%%%%%%%%%%%%%%%%%%%%%%%%%%%%%%%%%%%%%%%%%%%%%%%%%%%%%%%%%%%%%%%%%%%%%%%%%%%%%%%%%%%
%\twocolumngrid

%%%%%%%%%%%%%%%%%%%%%%%%%%%%%%%%%%%%%%%%%%%%%%%%%%%%%%%%%%%%%%%%%%%%%%%%%%%%%%%%%%%%%%%%%%%%%%%%%%%
\begin{figure}[!h]
	\centerline{\includegraphics[width=8.6cm,keepaspectratio]{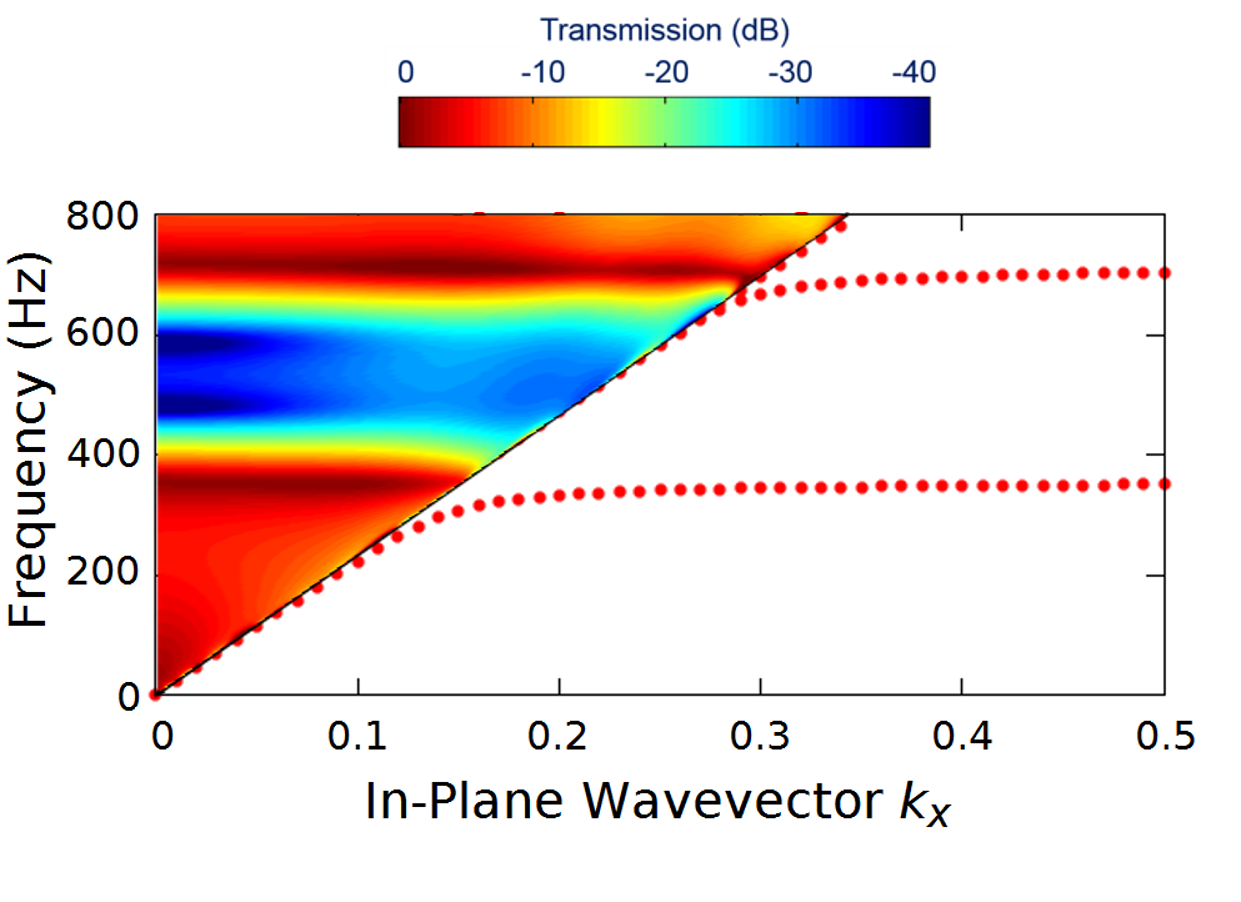}}
	\caption{Amplitude transmission spectra function of the In-Plane Wavevector $k_{x} = k_{0} sin(\theta_{i})$, and dispersion obtained numerically by finite element method. The acoustic metamaterial constituted by an array of resonators, where each unit cell is made of two Fabry-Perot cavities, one of which is space-coiled.  The metamaterial thickness is equal to $t=50$ mm, the system period $a=129$ mm, and the aperture width is $d=7$ mm.}
	\label{fig:fig5b}
\end{figure}
%%%%%%%%%%%%%%%%%%%%%%%%%%%%%%%%%%%%%%%%%%%%%%%%%%%%%%%%%%%%%%%%%%%%%%%%%%%%%%%%%%%%%%%%%%%%%%%%%%%

Figure~\ref{fig:fig5b} shows the dispersion calculated up to the frequency of $800$ Hz, for the unit cell shown in Figure~\ref{fig:fig00}a. The grating made of coupled cavities exhibits two modes, which are also materialized by two acoustic flat bands. Similarly to the case of a non-coupled space-coiled cavity grating, these modes appear to be spacially confined in the slit cavities. In the radiative zone, the coupling leads to a broad opacity frequency band which is nearly independent from the In-Plane Wavevector $k_{x}$, and consequently also insensitive to oblique incidence. Indeed, the interaction hereby presented is based on a sub-wavelength mode of structuring that works at a low frequency regime, well below the frequency of diffraction occurrence.

%%%%%%%%%%%%%%%%%%%%%%%%%%%%%%%%%%%%%%%%%%%%%%%%%%%%%%%%%%%%%%%%%%%%%%%%%%%%%%%%%%%

In what follows, we experimentally study the acoustic response of the previously presented metamaterial, using a four-microphone standing wave tube, more commonly referred to as a Kundt's tube. The sample is “sandwiched” between two parts of the tube, as shown in Fig. 9. For this purpose, the metamaterial was made of wood, as can be seen in Figure~\ref{fig:fig00}b, having the external dimensions of the square section Kundt's tube. The inner dimension of the tube square section is 140 mm, which limits the measurement to $900$ Hz. A loudspeaker, at one end of the tube, is used for generating a broadband random signal over the frequency range $50:900$ Hz. Acoustic pressure measurements are carried out for two different tube termination conditions, open and approximately anechoic. For each termination condition, complex pressure and velocity fields are estimated at each side of the sample material. These values are then used to calculate a two-by-two transfer matrix that characterizes the transmission through the sample~\cite{Olivieri_measurement_2006}. 

The acoustic metamaterial unit cell consists of a combination of two Fabry-Perot cavities placed side by side, as discussed above. The dimensions of the unit cell are the same as those used for simulation: $t=50$ mm, $a=129$ mm, $d=7$ mm, and thickness of wall $4$ mm. 
Figure~\ref{fig:fig8} represents the transmission spectrum of the sample under consideration. It clearly reveals an attenuation frequency band standing between two transmission peaks at $352$ Hz and beyond $744$ Hz. The attenuation band centered on $500$ Hz, and an attenuation of $28$ dB with a relative bandwidth of $33.3\%$. The maximum of attenuation reached is $37.5$ dB at the frequency of $434$ Hz. At this frequency, the antiresonance of the first mode, previously marked by letter E in Figure~\ref{fig:fig8}, is well established, as well as the second anti-resonance that occurs at $560$ Hz reaching $30$ dB of attenuation.
It is noteworthy that the wavelength is nearly $15$ times larger than the thickness structure, at the central frequency of $500$ Hz. The wavelength $\lambda$ is also nearly $5$ times larger than the array periodicity. 

When dealing with such subwavelength structures, it was notably showed that dissipation due to boundary layer effects cannot be neglected~\cite{ward_boundary-layer_2015}. In our numerical calculation of transmission spectrum, represented in Figure~\ref{fig:fig8}, we have included thermoacoustic equations in order to take into account dissipations. We barely attain $-3$ dB and $-4$ dB at the resonance frequencies, whereas numerical calculations without losses forecast a near-total transmission at the resonance. But more significantly, compared to experimental results, the attenuation band perfectly match, despite some fluctuations that can be seen at the second mode around $650$ Hz. Some discrepancies however exist between experiment and simulation. But some elements may help to explain these differences. Firstly, due to means of measurement, we fabricated a sample that fits with the waveguide shape. In doing so, the sample is therefore constituted of $1$ unit, considering the period $a=129$ mm previously mentioned. Moreover, some minor fabrication errors can also be mentioned. Such elements can affect the resonance frequencies of the cavities, as well as their quality factor, which can slightly disturb the coupling between the resonators. Nevertheless, the effect of low frequency sound screening is achieved, based on a phenomenon of Fano interaction realized with a cavity structure with a thickness that does not exceed $ \lambda /15 = 50$ mm.

%\section{Conclusions} 
\

In conclusion, we have proposed an acoustic metamaterial based on coupled resonant cavities demonstrating low frequency sound screening. The unit cell consists of two Fabry-Perot cavities, with an acoustic response that can be easily tailored through geometrical parameters. We experimentally obtained the acoustic transmission properties of this metamaterial, by using a Kundt's tube. As a result, the coupling between detuned resonators having different quality factors leads to asymmetric shape of peaks in the transmission spectra. Such asymmetric shapes of resonances and antiresonances, that are close together, permit to create an area of huge acoustic blocking effect.
We reach an attenuation of few tens of dB at low frequency, with a metamaterial thickness which is fifteen times smaller than the wavelength. In doing so, we avoid one of the main limitations of omnidirectional capabilities involving locally resonant perforated plates, which is namely diffraction. 
This has potential for many applications, particularly in sound insulation. More than that, the free choice and combination of the elementary bricks constituting the acoustic metamaterials represents a real opportunity to develop acoustic metasurfaces. Such devices, having the ability to bend a beam trajectory or even cloak objects \cite{wavemot2011}, would certainly increase our control over the propagation of waves.

\acknowledgments
The authors gratefully acknowledge Youssef Tejda for the fabrication of the device described in this work. The research leading to these results has received funding from the Region of Franche-Comte and financial support from the Labex ACTION program (Contact No. ANR-11-LABX-0001-01).


\begin{thebibliography}{0}

%

\bibitem{cummer_controlling_2016}
  \Name{ Cummer S. A., Christensen J. \and Al\'{u} A.}
  \REVIEW{Nature Reviews Materials}{1}{2016}{16001}.
% 

\bibitem{lee_acoustic_2017}
\Name{Lee D., Nguyen D. M. \and Rho J.}
\REVIEW{Nano Convergence}{4}{2017}{3}.

\bibitem{khelif_acoustic_2010}
\Name{Khelif A., Mohammadi S., Eftekhar A. A., Adibi A. \and Aoubiza B.}
\REVIEW{Journal of Applied Physics}{108}{2010}{084515}.

\bibitem{wilm_out--plane_2003}
\Name{Wilm M., Khelif A., Ballandras S., Laude V., \and Djafari-Rouhani B.}
\REVIEW{Physical Review E}{67}{2003}{065602}.

\bibitem{martinezN1995}
\Name{R. Mart\'{i}nez-Sala, J. Sancho, J. V. Sanchez, V. Gomez, J. Llinares, and F. Meseguer}
\REVIEW{Nature}{378}{1995}{241}.

\bibitem{khelif_locally_2010}
\Name{Khelif A., Achaoui Y., Benchabane S., Laude V. \and Aoubiza B.}
\REVIEW{Physical Review B}{81}{2010}{214303}.

\bibitem{addouche_superlensing_2014}
\Name{Addouche M., Al-Lethawe M. A., Choujaa A. \and Khelif A.}
\REVIEW{Applied Physics Letters}{105}{2014}{023501}.


\bibitem{LiuScience2000}
\Name{Liu Z., Zhang X., Mao Y., Zhu Y. Y., Yang Z., Chan C. T., \and Sheng P.}
\REVIEW{Science}{289}{2000}{1734}. 

\bibitem{yang_membrane-type_2008}
\Name{Yang Z., Mei J., Yang M., Chan N. H., \and Sheng P.}
\REVIEW{Phys. Rev. Lett}{101}{2008}{204301}.  


\bibitem{liang_extreme_2012}
\Name{Liang Z. \and Li J.}
\REVIEW{Physical Review Letters}{108}{2012}{114301}.  

\bibitem{frenzel_three-dimensional_2013}
\Name{Frenzel T., Brehm J. D., B\"{u}ckmann T., Schittny R., Kadic M., \and Wegener M.}
\REVIEW{Applied Physics Letters}{103}{2013}{061907}.
%

\bibitem{liang_space-coiling_2013}
\Name{Liang Z., Feng T., Lok S., Liu  F., Ng K. B., Chan C. H., Wang J., Han S., Lee S., \and Li J.}
\REVIEW{Scientific Reports}{3}{2013}{}.  

\bibitem{xie_measurement_2013}
\Name{Xie Y., Popa B.-I., Zigoneanu L., \and Cummer S. A.}
\REVIEW{Physical Review Letters}{110}{2013}{175501}.    


\bibitem{xie_tapered_2013}
\Name{Xie Y., Konneker A., Popa B.-I., \and Cummer S. A.}
\REVIEW{Applied Physics Letters}{103}{2013}{201906}.    

\bibitem{song_emission_2014}
\Name{Song K., Lee S.-H., Kim K., Hur S., \and J. Kim J.}
\REVIEW{Scientific Reports}{4}{2014}{}.    

\bibitem{li_extraordinary_2013}
\Name{ Li Y., Liang B., Zou X.-y., \and Cheng J.-c.}
\REVIEW{Applied Physics Letters}{103}{2013}{063509}.    

\bibitem{park_giant_2013}
\Name{Park J. J., Lee K. J. B., Wright O. B., Jung M. K., \and Lee S. H.}
\REVIEW{Physical Review Letters}{110}{2013}{244302}.    

\bibitem{lu_extraordinary_2007}
\Name{Lu M.-H., Liu X.-K., Feng L., Li J., Huang C.-P., Chen Y.-F., Zhu Y.-Y., Zhu S.-N., \and Ming N.-B.}
\REVIEW{Physical Review Letters}{99}{2007}{174301}.    

\bibitem{miroshnichenko_fano_2010}
\Name{Miroshnichenko A. E., Flach S., \and Kivshar Y. S.}
\REVIEW{Reviews of Modern Physics}{82}{2010}{2257}.    

\bibitem{Fano61}
\Name{Fano U.}
\REVIEW{Phys. Rev.}{124}{1961}{1866}.    

\bibitem{Christian2013}
\Name{Ott C., Kaldun A., Raith P., Meyer K., Laux M., Evers J., Keitel C. H., Greene C. H., \and Pfeifer T.}
\REVIEW{Science}{340}{2013}{716}.    

\bibitem{liu_plasmonic_2009}
\Name{Liu N., Langguth L., Weiss T., Kastel J., Fleischhauer M., Pfau T., \and Giessen H.}
\REVIEW{Nature Materials}{8}{2009}{758}. 
%\¨{a}   

\bibitem{Liu10}
\Name{Liu F., Ke M., Zhang A., Wen W., Shi J., Liu Z., \and Sheng P.}
\REVIEW{Phys. Rev. E}{82}{2010}{026601}.    

\bibitem{Santillan11}
\Name{Santill\'{a}n A. \and Bozhevolnyi  S. I.}
\REVIEW{Phys. Rev. B}{84}{2011}{064304}.    

\bibitem{amin_acoustically_2015}
\Name{Amin M., Elayouch A., Farhat M., Addouche M., Khelif A., \and Ba\u{g}c{\i} H.}
\REVIEW{Journal of Applied Physics}{118}{2015}{164901}. 

\bibitem{fleischhauer_electromagnetically_2005}
\Name{M. Fleischhauer, A. Imamoglu, and J. P. Marangos}
\REVIEW{Reviews of Modern Physics}{77}{2005}{633}. 

\bibitem{elayouch_experimental_2013}
\Name{Elayouch A., Addouche M., Herth E., \and Khelif A.}
\REVIEW{Applied Physics Letters}{103}{2013}{083504}. 

%%%%%%%%%%%%

\bibitem{Diffwiley2007}
  \Name{Ufimtsev P. Y. }
  \Book{Fundamentals of the Physical Theory of Diffraction}
  \Editor{Wiley}
  \Publ{Wiley, New York}
  \Year{2007}




\bibitem{elayouch_how_2016}
\Name{Elayouch A., Addouche M., Lasaygues P., Achaoui Y., Ouisse M., \and Khelif A.}
\REVIEW{Comptes Rendus Physique Phononic crystals / Cristaux phononiques}{17}{2016}{518}. 

\bibitem{royerBOOK1999}
\Name{Royer D. \and Dieulesaint  E.}
\Book{Elastic waves in solids}
\Editor{Wiley}
\Publ{Wiley, New York}
\Year{1999}


\bibitem{_comsol_2008}
\Name{COMSOL Multiphysics Version 3.5a}
\Book{User's guide COMSOL MULTIPHYSICS}
\Editor{Comsol}
\Year{2008}




\bibitem{khelifBOOK2015}
\Name{Khelif A. \and  Adibi A.}
\Book{ Phononic Crystals}
\Editor{Wiley}
\Publ{Springer-Verlag New York Inc., New York}
\Year{2015}


\bibitem{kinslerBOOK1999}
\Name{Kinsler L. E. , Frey A. R., Coppens A. B., \and Sanders J. V.}
\Book{Fundamentals of Acoustics}
\Editor{Wiley}
\Publ{Springer-Verlag New York Inc., New York}
\Year{1999}


\bibitem{Gal2:2011}
\Name{Gallinet B. \and Martin O. J.}
\REVIEW{ACS Nano}{ 5}{2011}{8999}. 

\bibitem{miroshnichenko_fano_2010}
\Name{Miroshnichenko A. E., Flach S., \and Kivshar Y. S.}
\REVIEW{Reviews of Modern Physics}{82}{2010}{2257}. 

\bibitem{Olivieri_measurement_2006}
\Name{Olivieri O., Bolton J. S., \and Yoo T.}
\REVIEW{Proc. 35th International Congress and Exposition on Noise Control Engineering (INTER-NOISE 2006)}{8}{2006}{}. 

\bibitem{ward_boundary-layer_2015}
\Name{Ward G., Lovelock R., Murray A., Hibbins A., Sambles J., \and Smith J.}
\REVIEW{Physical Review Letters}{115}{2015}{044302}. 

\bibitem{wavemot2011}
\Name{Dupont G., Farhat M., Diatta A., Guenneau S., \and Enoch S.}
\REVIEW{Wave Motion}{48}{2011}{483-496}.


\end{thebibliography}
\end{document}